\documentclass[fleqn,usenatbib]{mnras}

\usepackage{amsmath}	

\usepackage{graphicx, amssymb, aas_macros, natbib, bm}
\usepackage{marvosym}
\usepackage{enumitem}
\usepackage{fancybox}
\usepackage{tabularx}
\usepackage{tabulary}
\newcommand{\vmax}{{\it V}_{\rm{max}}}

\newcommand{\mvir}{M_{\rm{vir}}}
\newcommand{\mhalo}{{M}_{\rm{halo}}}
\newcommand{\mpeak}{{M}_{\rm{peak}}}
\newcommand{\rvir}{R_{\rm vir}}
\newcommand{\rscale}{R_{\rm s}}

\newcommand{\vpeak}{{\it V}_{\rm{peak}}}

\newcommand{\mstar}{{M}_{\star}}

\newcommand{\msun}{{\rm M}_{\odot}}

\newcommand{\gyr}{{\rm Gyr}}
\newcommand{\kpc}{{\rm kpc}}

\newcommand{\lt}{<}

\usepackage[T1]{fontenc}
\usepackage{ae,aecompl}

\usepackage{newtxtext,newtxmath}

\title[Low-Mass Satellite Quenching with {\it Gaia}]
{Characterizing the Infall Times and Quenching Timescales of Milky Way
  Satellites with \emph{Gaia} Proper Motions}

\author[Fillingham et al.]
{Sean P.~Fillingham,$^1$\thanks{$\!\!$e-mail: sfilling@uci.edu}\ 
  M.~C.~Cooper,$^1$ 
Tyler Kelley,$^1$
M.~K.~Rodriguez Wimberly,$^1$
\newauthor
Michael Boylan-Kolchin,$^2$
James S.~Bullock,$^1$
Shea Garrison-Kimmel,$^3$
\newauthor
Marcel S. Pawlowski,$^{4, 1}$\thanks{$\!\!$Schwarzschild Fellow}\thanks{$\!\!$Hubble Fellow}
Coral Wheeler $^3$ \\
$\!\!^1$Center for Cosmology, Department of Physics \& Astronomy,
University of California, Irvine, 4129 Reines Hall, Irvine, CA 92697, USA \\
$\!\!^2$Department of Astronomy, The University of Texas at Austin,
2515 Speedway, Stop C1400, Austin, TX 78712, USA \\
$\!\!^3$TAPIR, Mailcode 350-17, California Institute of Technology,
Pasadena, CA 91125, USA \\
$\!\!^4$Leibniz-Institut f\"{u}r Astrophysik Potsdam (AIP), An der Sternwarte 16, D-14482 Potsdam, Germany
}

\begin{document}
\pagerange{\pageref{firstpage}--\pageref{lastpage}} 
\pubyear{2019}
\maketitle

\label{firstpage}
\begin{abstract}
  Observations of low-mass satellite galaxies in the nearby Universe point
  towards a strong dichotomy in their star-forming properties relative to
  systems with similar mass in the field. Specifically, satellite galaxies are
  preferentially gas poor and no longer forming stars, while their field
  counterparts are largely gas rich and actively forming stars. Much of the
  recent work to understand this dichotomy has been statistical in nature,
  determining not just that environmental processes are most likely responsible
  for quenching these low-mass systems but also that they must operate very
  quickly after infall onto the host system, with quenching timescales
  $\lesssim 2~ {\rm Gyr}$ at $\mstar \lesssim 10^{8}~\msun$.
  This work utilizes the newly-available \emph{Gaia} DR2 proper motion
  measurements along with the Phat ELVIS suite of high-resolution,
  cosmological, zoom-in simulations to study low-mass satellite quenching
  around the Milky Way on an object-by-object basis.
  We derive constraints on the infall times for $37$ of
  the known low-mass satellite galaxies of the Milky Way, finding that
  $\gtrsim~70\%$ of the ``classical'' satellites of the Milky Way are consistent
  with the very short quenching timescales inferred from the total population in
  previous works. The remaining classical Milky Way satellites have quenching
  timescales noticeably longer, with $\tau_{\rm quench} \sim 6 - 8~{\rm Gyr}$,
  highlighting how detailed orbital modeling is likely necessary to understand the
  specifics of environmental quenching for individual satellite
  galaxies. Additionally, we find that the $6$ ultra-faint dwarf galaxies with
  publicly available {\it HST}-based star-formation histories are all consistent
  with having their star formation shut down prior to infall onto the Milky Way
  -- which, combined with their very early quenching times, strongly favors
  quenching driven by reionization. 

\end{abstract}


\section{Introduction}
\label{sec:intro}

Large surveys of the nearby Universe demonstrate that galaxies at (and below) the mass
scale of the Magellanic clouds
($10^{6}~\msun \lesssim \mstar \lesssim 10^{9}~\msun$) are predominantly star
forming in the field \citep{haines08, geha12}.
This stands in stark contrast to the population of low-mass galaxies that
currently reside near a more massive host system, where the fraction of systems
that are no longer forming stars (i.e.~``quenched'') is significantly larger 
\citep{weinmann06, geha12, phillips15a}.
For dwarfs with $\mstar \lesssim 10^{8}~\msun$ in the Local Volume, this
field-satellite dichotomy is very apparent, with satellites of the Milky Way
(MW) and M31 being largely gas-poor, passive systems in contrast to the
gas-rich, star-forming field population \citep[e.g.][]{mateo98, grcevich09,
  spekkens14}.
This clear distinction between the field and satellite populations strongly
favors environmental processes as the dominant quenching mechanisms in this
low-mass regime \citep[$\mstar \lesssim 10^{8}~\msun$,][]{lin83, slater14,
  weisz15, wetzel15b, fham15, fham16, fham18, simpson18}.
At the very lowest mass scales (i.e.~the regime of ultra-faint dwarfs), however,
there is evidence for a transition in the dominant quenching mechanism from one
associated with galaxy environment to one driven by reionization. 
%
The universally old stellar populations observed in the ultra-faint dwarfs
(UFDs) suggest that star formation on the smallest galactic scales is suppressed
at very early times, favoring a transition in the dominant quenching
mechanism at $\mstar \lesssim 10^{5}~\msun$
\citep{brown14, wimberly19}.

Across all mass scales, some of the most powerful studies of satellite quenching
have utilized measurements of satellite and field quenched fractions to infer
the timescale upon which satellite quenching occurs following infall
\citep[e.g.][]{wetzel13, fham15, balogh16, fossati17}.
These studies point to a picture where quenching proceeds relatively slowly at
high satellite masses, consistent with quenching via starvation
\citep{wheeler14, fham15, davies16, trussler18}.
Below some host-dependent critical mass scale, however,
quenching is rapid, as stripping becomes increasingly efficient \citep{fham16}.
In defining this model of satellite quenching, simulations are commonly utilized
to constrain the distribution of infall times for an observed sample of
satellites. This statistical approach is required, as it is extremely difficult
to infer the infall time for a significant fraction of the satellite population
in even the most nearby groups and clusters.
Moreover, in systems more distant than $\sim1$~Mpc, it is difficult to measure a
precise star-formation history via spatially-resolved stellar photometry, even
with the aid of imaging from the {\it Hubble Space Telescope} ({\it HST}).
Within the Local Group, however, we are afforded the luxury of more detailed observations
of the nearby satellite and field populations. This is particularly true with
the release of {\it Gaia} Data Release 2 \citep[DR2,][]{gaia, gaiaDR2}, which
now enables an investigation of satellite quenching timescales (measured
relative to infall) on an object-by-object basis.
This offers a unique opportunity to test the results of large statistical
analyses and our current physical picture of satellite quenching.

In this work, we aim to determine the quenching timescale and ultimately
constrain the potential mechanisms responsible for suppressing star formation in
individual MW satellite galaxies.
Utilizing the latest data products from {\it Gaia} DR2 \citep{gaiaDR2,
  gaiaDR2PM}, we infer the cosmic time when each dwarf galaxy around the MW
became a satellite (i.e.~the infall time) through comparison to the
Phat ELVIS suite of cosmological zoom-in simulations of MW size
galaxies \citep{kelley18}.
In addition, we infer the quenching times for the MW satellites based on their
published star-formation histories, as derived from {\it HST} imaging
\citep{weisz14a, weisz15, brown14}. Finally, through comparison of the quenching
times to the infall times, we characterize the quenching timescales for each
object and constrain the potential mechanisms responsible for quenching each MW
satellite galaxy.
In \S\ref{sec:MWdwarfs}, we discuss our sample of local dwarfs and the
methodology by which we measure the infall and quenching time for each system.
Our primary results are presented in \S\ref{sec:results}, followed by a
discussion of how these results connect to physical models of satellite
quenching in \S\ref{sec:disc}. Finally, we summarize our results and conclusions
in \S\ref{sec:summary}.
Where necessary, we adopt a $\Lambda$CDM cosmology with the following
parameters: $\sigma_8 = 0.815$, $\Omega_{m} =
0.3121$, $\Omega_{\Lambda} = 0.6879$, $n_{s} = 0.9653$, and $h =
0.6751$ \citep{planck16}, consistent with the simulations used in this
work.


\begin{table*}
  \centering
  \setlength{\tabcolsep}{12pt}
  \def\arraystretch{1.35}
  \begin{tabular}{lccccccc}
    \hline\hline
    $\rm Name$ & $\rm Distance$ & $V_{\rm 3D}$ & $V_{\rm r}$ &$V_{\rm
                                                               tan}$ &
                                                                       $\it e$ & $t_{90}$ & $t_{\rm infall}$ \\

    & (kpc) & (km/s) & (km/s) & (km/s) &   & (Gyr) & (Gyr)\\
    & (1) & (2) & (3) & (4) & (5) & (6) & (7)\\
    
  \hline
  \hline
Sag I & $18$ & $312^{+21}_{-18}$ & $142~\pm~1$ & $278^{+23}_{-20}$ & $0.42^{+0.03}_{-0.02}$ & $3.4^{+5.2}_{-3.1}$ & $10.6^{+1.6}_{-1.9}$ \\ 
Tuc III & $21$ & $-236^{+5}_{-5}$ & $-228~\pm~2$ & $60^{+12}_{-12}$ & $0.86^{+0.03}_{-0.03}$ & $ -- $ & $9.5^{+2.4}_{-2.8}$ \\ 
Dra II & $23$ & $-355^{+25}_{-24}$ & $-155~\pm~8$ & $319^{+27}_{-27}$ & $0.53^{+0.07}_{-0.06}$ & $ -- $ & $10.2^{+1.8}_{-2.4}$ \\ 
Hyd I & $25$ & $-370^{+14}_{-13}$ & $-57~\pm~2$ & $365^{+14}_{-13}$ & $0.49^{+0.06}_{-0.05}$ & $ -- $ & $9.4^{+1.7}_{-1.8}$ \\ 
Seg 1 & $27$ & $232^{+28}_{-26}$ & $116~\pm~4$ & $200^{+31}_{-30}$ & $0.39^{+0.06}_{-0.04}$ & $ -- $ & $10.8^{+1.3}_{-1.4}$ \\ 
Car III & $28$ & $387^{+33}_{-30}$ & $45~\pm~4$ & $384^{+33}_{-30}$ & $0.58^{+0.12}_{-0.11}$ & $ -- $ & $7.6^{+3.4}_{-2.7}$ \\ 
Ret II & $32$ & $-248^{+15}_{-14}$ & $-102~\pm~2$ & $226^{+17}_{-16}$ & $0.31^{+0.02}_{-0.02}$ & $ -- $ & $10.2^{+1.1}_{-2.4}$ \\ 
Tri II & $34$ & $-333^{+20}_{-18}$ & $-255~\pm~3$ & $213^{+29}_{-28}$ & $0.71^{+0.02}_{-0.02}$ & $ -- $ & $9.5^{+1.5}_{-2.1}$ \\ 
Car II & $37$ & $355^{+16}_{-14}$ & $203~\pm~3$ & $291^{+19}_{-18}$ & $0.64^{+0.04}_{-0.03}$ & $ -- $ & $7.9^{+2.5}_{-2.4}$ \\ 
Boo II & $39$ & $-383^{+76}_{-68}$ & $-54~\pm~9$ & $378^{+79}_{-70}$ & $0.63^{+0.28}_{-0.27}$ & $ -- $ & $1.1^{+0.6}_{-0.6}$ \\ 
U Maj II & $40$ & $-288^{+21}_{-19}$ & $-58~\pm~2$ & $282^{+21}_{-19}$ & $0.26^{+0.1}_{-0.05}$ & $ -- $ & $10.7^{+1.4}_{-2.3}$ \\ 
Seg 2 & $42$ & $224^{+39}_{-34}$ & $73~\pm~3$ & $211^{+40}_{-35}$ & $0.27^{+0.09}_{-0.04}$ & $ -- $ & $10.8^{+1.6}_{-1.9}$ \\ 
Wil 1 & $42$ & $120^{+56}_{-44}$ & $23~\pm~4$ & $118^{+57}_{-47}$ & $0.54^{+0.21}_{-0.24}$ & $ -- $ & $10.7^{+1.3}_{-1.4}$ \\ 
ComBer I & $43$ & $276^{+30}_{-27}$ & $31~\pm~3$ & $274^{+31}_{-28}$ & $0.2^{+0.13}_{-0.06}$ & $13.0^{+14.3}_{-11.8}$ & $10.2^{+2.6}_{-3.3}$ \\ 
Tuc II & $53$ & $-283^{+24}_{-20}$ & $-187~\pm~2$ & $212^{+33}_{-29}$ & $0.58^{+0.03}_{-0.02}$ & $ -- $ & $9.5^{+2.0}_{-1.9}$ \\ 
Boo I & $63$ & $192^{+27}_{-25}$ & $94~\pm~2$ & $167^{+32}_{-31}$ & $0.41^{+0.12}_{-0.08}$ & $12.6^{+13.7}_{-11.6}$ & $10.7^{+0.6}_{-1.9}$ \\ 
U Min I & $77$ & $-153^{+17}_{-16}$ & $-71~\pm~2$ & $135^{+19}_{-19}$ & $0.49^{+0.07}_{-0.08}$ & $10.2^{+11.7}_{-7.7}$ & $10.7^{+1.7}_{-2.0}$ \\ 
Dra I & $79$ & $-160^{+19}_{-15}$ & $-88~\pm~2$ & $133^{+22}_{-19}$ & $0.53^{+0.07}_{-0.09}$ & $9.1^{+10.7}_{-5.8}$ & $10.4^{+2.4}_{-3.1}$ \\ 
Hor I & $83$ & $-213^{+48}_{-44}$ & $-34~\pm~3$ & $210^{+49}_{-44}$ & $0.21^{+0.18}_{-0.08}$ & $ -- $ & $8.8^{+1.8}_{-2.0}$ \\ 
Scu I & $84$ & $198^{+21}_{-22}$ & $74~\pm~1$ & $184^{+22}_{-24}$ & $0.32^{+0.07}_{-0.04}$ & $10.6^{+11.9}_{-7.1}$ & $9.9^{+1.7}_{-2.9}$ \\ 
Sext I & $88$ & $241^{+25}_{-22}$ & $79~\pm~1$ & $228^{+27}_{-23}$ & $0.3^{+0.07}_{-0.02}$ & $ -- $ & $8.4^{+2.7}_{-0.9}$ \\ 
U Maj I & $101$ & $257^{+62}_{-53}$ & $9~\pm~2$ & $257^{+62}_{-53}$ & $0.31^{+0.34}_{-0.22}$ & $11.2^{+12.5}_{-10.0}$ & $1.5^{+5.1}_{-1.6}$ \\ 
Aqu II & $105$ & $250^{+241}_{-164}$ & $49~\pm~8$ & $244^{+242}_{-174}$ & $0.75^{+0.24}_{-0.45}$ & $ -- $ & $1.6^{+5.4}_{-3.5}$ \\ 
Car I & $105$ & $162^{+21}_{-22}$ & $1~\pm~2$ & $162^{+21}_{-22}$ & $0.27^{+0.12}_{-0.12}$ & $2.2^{+3.7}_{-2.2}$ & $9.9^{+0.6}_{-2.7}$ \\ 
Cra II & $111$ & $-113^{+24}_{-19}$ & $-83~\pm~2$ & $75^{+34}_{-35}$ & $0.74^{+0.13}_{-0.15}$ & $ -- $ & $7.8^{+2.7}_{-2.0}$ \\ 
Gru I & $116$ & $-274^{+102}_{-69}$ & $-195~\pm~4$ & $190^{+127}_{-117}$ & $0.81^{+0.17}_{-0.11}$ & $ -- $ & $1.1^{+1.0}_{-0.9}$ \\ 
Her I & $128$ & $163^{+31}_{-9}$ & $152~\pm~1$ & $59^{+61}_{-37}$ & $0.85^{+0.1}_{-0.18}$ & $11.8^{+13.2}_{-10.5}$ & $6.6^{+2.3}_{-0.7}$ \\ 
Frn I & $141$ & $-137^{+26}_{-25}$ & $-41~\pm~1$ & $131^{+27}_{-27}$ & $0.42^{+0.14}_{-0.13}$ & $ -- $ & $10.7^{+0.8}_{-3.1}$ \\ 
Cra I & $144$ & $-112^{+133}_{-77}$ & $-10~\pm~3$ & $111^{+132}_{-79}$ & $0.63^{+0.31}_{-0.38}$ & $ -- $ & $8.9^{+3.0}_{-3.1}$ \\ 
Hyd II & $147$ & $284^{+259}_{-141}$ & $118~\pm~6$ & $257^{+272}_{-177}$ & $0.89^{+0.1}_{-0.36}$ & $2.2^{+2.5}_{-2.0}$ & $10.5^{+1.5}_{-2.4}$ \\ 
Leo IV & $154$ & $312^{+306}_{-217}$ & $1~\pm~8$ & $311^{+306}_{-217}$ & $0.95^{+0.05}_{-0.62}$ & $12.2^{+13.6}_{-10.7}$ & $10.4^{+1.4}_{-1.4}$ \\ 
CVn II & $160$ & $-182^{+150}_{-77}$ & $-93~\pm~4$ & $157^{+163}_{-108}$ & $0.71^{+0.27}_{-0.26}$ & $12.7^{+14.3}_{-11.1}$ & $9.0^{+1.0}_{-2.8}$ \\ 
Leo V & $173$ & $311^{+307}_{-210}$ & $50~\pm~7$ & $308^{+309}_{-219}$ & $0.96^{+0.04}_{-0.58}$ & $ -- $ & $10.6^{+1.4}_{-1.5}$ \\ 
Pis II & $181$ & $-400^{+434}_{-265}$ & $-64~\pm~8$ & $394^{+438}_{-275}$ & $0.99^{+0.01}_{-0.48}$ & $ -- $ & $8.3^{+1.8}_{-1.8}$ \\ 
CVn I & $210$ & $125^{+67}_{-38}$ & $81~\pm~2$ & $94^{+78}_{-62}$ & $0.66^{+0.24}_{-0.25}$ & $8.3^{+9.5}_{-6.3}$ & $9.4^{+0.9}_{-2.3}$ \\ 
Leo II & $227$ & $76^{+77}_{-44}$ & $19~\pm~1$ & $74^{+78}_{-49}$ & $0.73^{+0.2}_{-0.44}$ & $6.4^{+7.2}_{-5.8}$ & $7.8^{+3.3}_{-2.0}$ \\ 
Leo I & $272$ & $181^{+44}_{-13}$ & $166~\pm~1$ & $71^{+79}_{-48}$ & $0.87^{+0.09}_{-0.09}$ & $1.7^{+1.9}_{-1.6}$ & $2.3^{+0.6}_{-0.5}$ \\ 

\hline

\label{table:MWdwarfs} 
\end{tabular} 
\caption{
  Properties of satellite galaxies of the Milky Way, as selected from 
  \citet{fritz18} and listed in order of increasing Galactocentric distance.
  (1) Distance from the center of the MW in kpc \citep{fritz18}.
  (2) Total velocity, (3) Radial velocity, and (4) Tangential
  velocity, all in ${\rm km}~{\rm s}^{-1}$, in the
  Galactocentric frame-of-reference \citep{fritz18}.
  (5)  Orbital eccentricity, $e$, based on the ``heavy'' MW mass used
  in \citet{fritz18}.
  (6) Quenching time ($t_{90}$), in Gyr, inferred from published SFHs by adopting the
  lookback time at which the dwarf galaxy formed $90\%$ of its current
  stellar mass \citep{weisz14a, brown14}. The uncertainty on $t_{90}$
  as given by the $1\sigma$ bounds on the quenching time, in Gyr, inferred
  from the SFHs by adopting the lookback-time at which the $1\sigma$
  bounds on the SFH crossed the $90\%$ mass threshold. 
  (7) Infall time, in Gyr, inferred from the peak in the infall time
  KDE from matching to subhalos in the pELVIS
  simulations (see~\S\ref{subsec:infalltime}). The uncertainty in the
  infall time, in Gyr, as given by the upper and lower bounds on the KDE
  that encompass $68\%$ of the distribution centered on the peak, or
  roughly $1\sigma$ (see the magenta shaded region in
  Figure~\ref{fig:example}). 
  This table will be available at https://sfillingham.github.io after the paper is
  accepted for publication. }

\end{table*} 


\section{Milky Way Dwarf Galaxies}
\label{sec:MWdwarfs}

As shown in Table~\ref{table:MWdwarfs}, our sample of Milky Way satellite
galaxies is selected from \citet{fritz18}, totalling $37$ systems located within
$300~{\rm kpc}$ of the MW and for which the {\it Gaia} DR2 dataset yields a
proper motion measurement.
Combined with measurements of distance and line-of-sight velocity from the
literature, the {\it Gaia} observations constrain the Galactocentric position
and $3{\rm D}$ velocity of each system (i.e.~full $6{\rm D}$ phase-space
information).
For those galaxies located within a Galactocentric distance of $300~{\rm kpc}$
(i.e.~within what we assume to be the virial radius of the MW), {\it
  HST}/WFCP2 and ACS imaging of $15$ systems yields estimates of their
star-formation history.
These $15$ dwarfs comprise our primary sample, spanning a stellar mass range of
$\mstar \sim 10^{3.5-8}~\msun$. All of the systems show no evidence of ongoing
star formation, with minimal H{\scriptsize I} reservoirs
\citep[$M_{\rm H\protect\scalebox{0.55}{\rm I}}~/~\mstar< 0.1$,][]{grcevich09,
  spekkens14}.
While this sample is undoubtedly incomplete below $\mstar \sim
10^{5.5}~\msun$ \citep{irwin94, tollerud08}, particularly at large
Galactocentric distances, our analysis focuses on the potential quenching
mechanism for each system individually, such that it does not rely strongly on
uniformly sampling the MW satellite population.
Another source of incompleteness is expected to be the limited sky
coverage of imaging surveys such as the Sloan Digital Sky Survey
\citep{york00} and the Dark Energy Survey \citep{DES16}, which should
not be significantly correlated with the accretion history of the
satellite population.

To better constrain the physical mechanisms responsible for quenching the
low-mass satellites of the Milky Way, we aim to characterize the quenching
timescale for each system.
As we define more explicitly below, the quenching timescale is the difference
between the infall time and the quenching time --- i.e.~the time that a
satellite spends inside the host dark matter halo, following infall, prior to
having its star formation shut down.
To infer the quenching timescale, we must first measure the quenching time and
infall time for each galaxy in our sample.


\begin{figure*}
 \centering
 \hspace*{-0.3in}
 \includegraphics[width=6in]{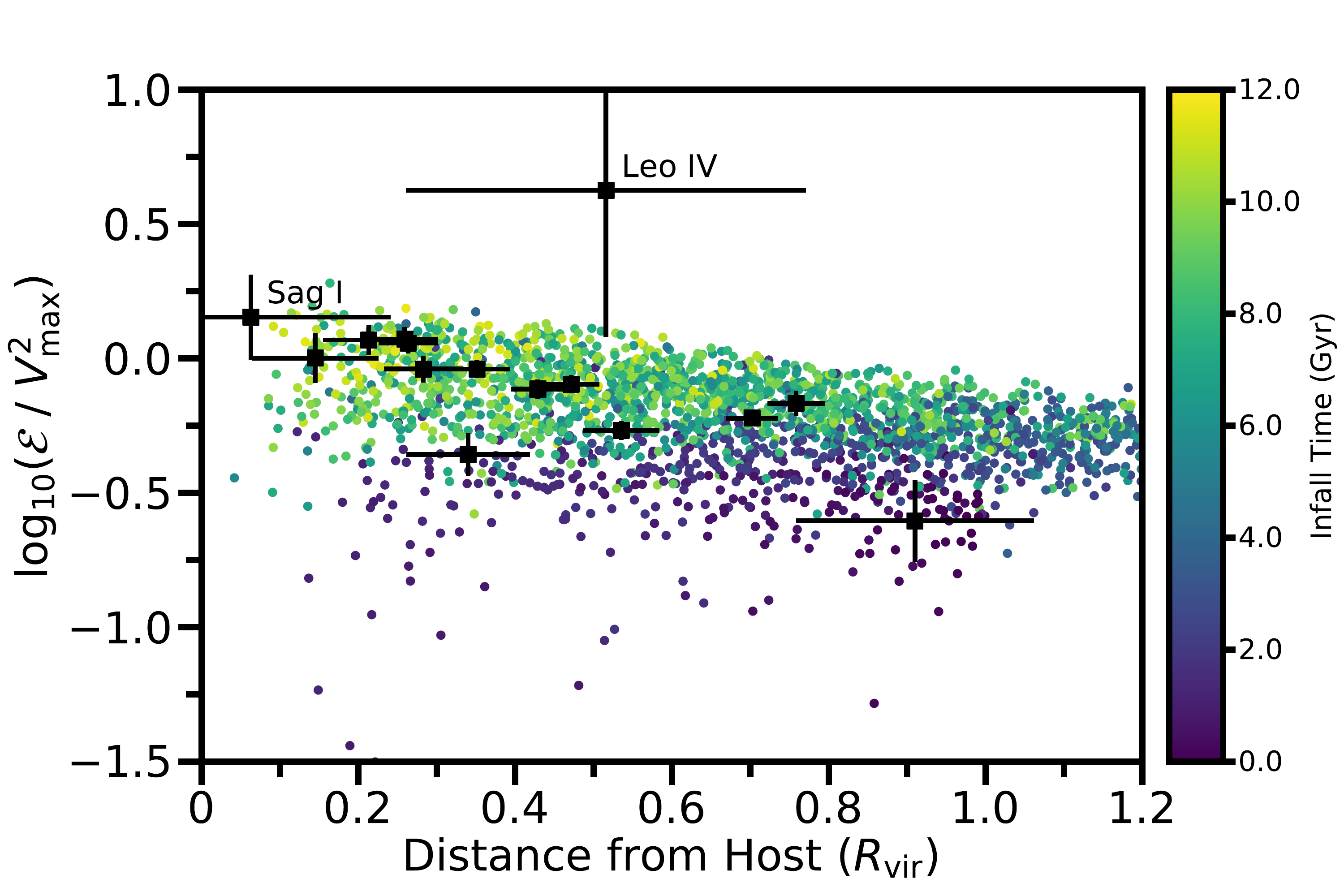}
 \caption{The binding energy ($\mathcal{E}$), scaled by the present-day $\vmax$
   of the host, as a function of distance from the host (in units of host
   $\rvir$) for subhalos in the pELVIS simulation suite. Each point is
   color-coded by the lookback time to first infall for the subhalo.  The black
   squares correspond to the Milky Way satellites in our primary sample, assuming
   $\mvir = 1.3 \times 10^{12}~\msun$, $\rvir = 300~{\rm kpc}$,
   $\rscale = 18.75~{\rm kpc}$, and $\vmax = 200~{\rm km/s}$. To constrain $t_{\rm infall}$, we select the
   $15$ subhalos closest to each MW dwarf in the plotted parameter space,
   restricting to only those subhalos that also have the same directionality in
   their host-centric radial velocity component. The error bars illustrate the
   selection region for each MW dwarf based on this selection criteria (see
   \S\ref{sec:MWdwarfs} for more details). There exists a strong gradient in
   typical infall time within this parameter space. Many of the Milky Way
   satellites reside in regions with well-defined infall times, such that strong
   constraints can be placed on $t_{\rm infall}$. Two objects, Sag~I
   and Leo~IV, are not well matched in this analysis, see
   \S\ref{sec:disc} for a more detailed discussion.}
 \label{fig:bedist}
\end{figure*}


\subsection{Quenching Time}

A galaxy's star-formation history (SFH) provides a direct constraint on the
epoch at which a galaxy quenched. 
For nearby systems, where the stellar population can be resolved and imaging can
reach the oldest main sequence turn-off population, we can constrain the SFH
with relative precision \citep{dolphin97, weisz11}.
Here, we adopt the lookback time at which a galaxy forms $90\%$ of its present
day stellar mass ($t_{90}$) to be the quenching time. 
Following \citet{weisz15}, we utilize $t_{90}$ (versus $t_{100}$) to minimize
the potential uncertainty associated with the modeling of blue straggler
populations.

For $15$ of the dwarfs in our sample (see Table~\ref{table:MWdwarfs}), there are
published star-formation histories from \citet{weisz14a} and \citet{brown14},
based on {\it HST}/WFPC2 and ACS imaging.
In the case of three systems (all UFDs), the SFH is measured by both
\citet{brown14} and \citet{weisz14a}, with excellent agreement between the two
datasets for Hercules and Leo~IV. For the third dwarf in common (CVn~II),
\citet{weisz14a} find a more extended SFH, in contrast to the largely ancient
stellar population inferred by \citet{brown14}.
Given that the {\it HST}/WPFC2 imaging analyzed by \citet{weisz14a} is
shallower and covers a smaller area than the {\it HST}/ACS imaging employed by
\citet{brown14}, we opt to utilize the SFH from \citet{brown14}. 
Even allowing for the more extended SFH measured by \citet{weisz14a}, our
results are qualitatively unchanged, such that we find CVn~II quenches prior to
infall.
It should be noted that recent simulations of dwarf galaxies have found that
stellar age gradients can exist and depend strongly on the SFH and merger
history of an individual dwarf galaxy \citep{elbadry16, graus19}.
These gradients can introduce an observational bias that varies depending on the
location of the \emph{HST} field relative to the galaxy half-light radius, with
the potential to shift the inferred median stellar age by as much as 
$\pm2~\gyr$ \citep{graus19}.
For the SFHs from \citet{weisz14a} and \citet{brown14}, the corresponding {\it
  HST} imaging fields are, on average, biased towards the center of each
system, so as to weakly favor younger SFHs and thus slightly
underestimate $t_{90}$.

To determine the time at which each system formed $90\%$ of its stellar mass
($t_{90}$), we linearly interpolate the measured SFHs from \citet{weisz14a},
using the published data tables \citep{weisz14aCATS}.
As shown in Figure~\ref{fig:example} for Leo~II, the corresponding uncertainty
in $t_{90}$ is determined by similarly interpolating and evaluating the SFHs
including their associated $1\sigma$ random and systematic errors.
%
%
For the remaining $6$ systems in our sample, we visually inspect the published
SFHs from \citet{brown14} to measure $t_{90}$. In particular, utilizing the
$1\sigma$ confidence intervals from Figure~8 of \citet{brown14}, we select the
lookback time at which the upper and lower bounds of the SFH reach $90\%$ of the
present day mass of the system. The average of these two times is taken to be
$t_{90}$. 
For Hercules and Leo~IV, we find a very minor difference between the $t_{90}$
inferred from using the \citet{weisz14a} dataset and from visually inspecting
the \citet{brown14} results ($\Delta t_{90} \lesssim 1~{\rm Gyr}$).
The $t_{90}$ measurements for each system in our sample, which we take to be the
quenching time ($t_{\rm quench}$), are listed in Table~\ref{table:MWdwarfs}
along with their corresponding uncertainties.

\subsection{Infall Time}
\label{subsec:infalltime}

For low-mass dwarfs in the Local Volume, there is a stark difference in the
star-forming properties of those within the virial radius ($\sim300~{\rm kpc}$)
of either the MW or M31 and those in the field \citep[e.g.][]{spekkens14}.
While it is uncertain as to the exact physical scale at which environmental
effects begin to affect satellite galaxies, this observed field-satellite
dichotomy makes a strong case that infall onto the host system (as defined by
crossing within the host's virial radius) marks the onset of environmental
quenching.
In order to constrain the timescale upon which environmental processes act
(i.e.~the quenching timescale), we must infer the infall time --- i.e.~the time
at which a satellite crosses within the virial radius of the host --- for each
dwarf in our sample.

\subsubsection{Gaia Proper Motions}
\label{subsubsec:gaia}

The {\it Gaia} mission recently transformed the observational landscape,
providing proper motion measurements for stars in a large number of local dwarfs
\citep{gaia, gaiaDR1, gaiaDR2}.
Using these data, several studies inferred proper motions for the Milky Way
satellites, utilizing various membership criteria \citep{gaiaDR2PM, simon18b,
  fritz18, nk18, massari18, pace19}.
Here, we use the results from \citet{fritz18}, as they provide the most
comprehensive set of proper motion measures for the MW satellite population.
For our primary sample, the observed proper motions as measured by
\citet{fritz18} agree (within the quoted errors) with those estimated by 
both \citet{simon18b} as well as \citet{gaiaDR2PM}.
From \citet{fritz18}, we adopt the Galactocentric distance, radial velocity,
tangential velocity, $3$D velocity, and eccentricity measures for each system
(see Table~\ref{table:MWdwarfs}).
Using this phase-space information, we match the MW satellites to a
subhalo population drawn from high-resolution, cosmological, zoom-in
simulations in an effort to constrain their infall times.
It should be noted that the proper motions (and resulting
Galactocentric velocities) in some cases have very large
uncertainties, which lead to an overestimation of the proper
motion.
We therefore expect that some of the objects in this analysis
(e.g. Leo IV) will have unusually large Galactocentric velocities.
Ultimately this uncertainty only affects Leo~IV in our primary sample (see
Fig.~\ref{fig:bedist}), and does not change our overall conclusions.

\begin{figure*}
 \centering
 \hspace*{-0.02in}
 \includegraphics[width=0.925\textwidth]{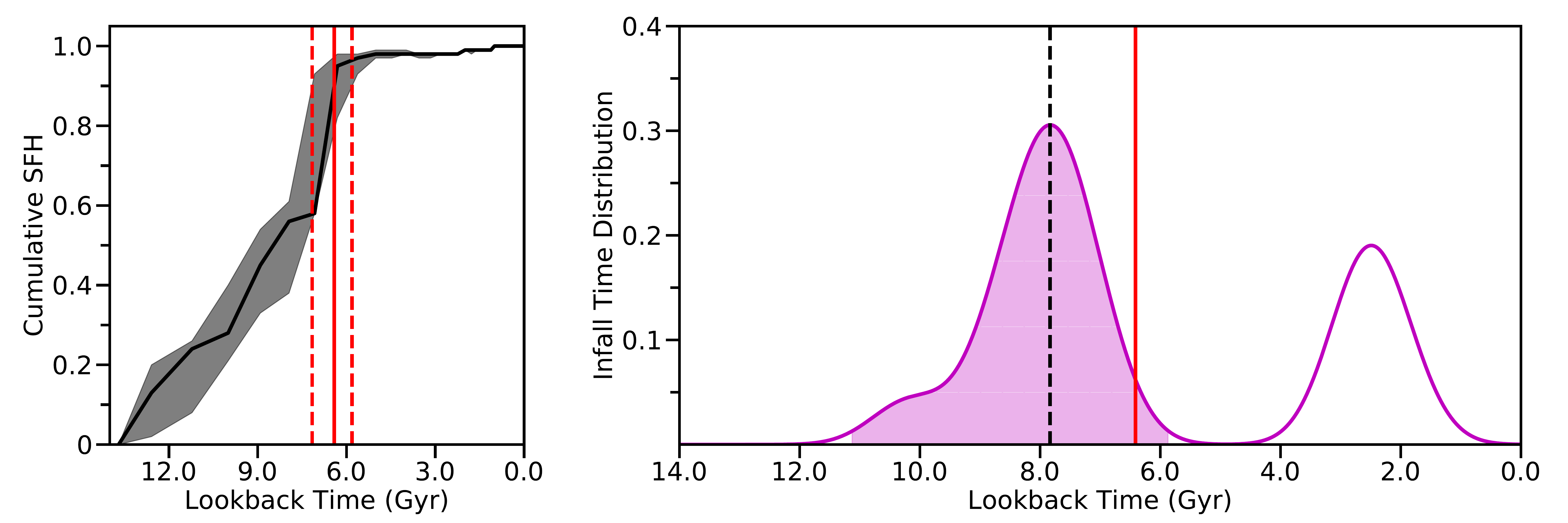}
 \caption{Illustrative example of both the quenching time and infall time
   measurements. \emph{Left}: The cumulative SFH for Leo~II, adapted from
   \citet{weisz14a}, showing the adopted quenching time (solid red line) and
   corresponding $1\sigma$ uncertainty in the quenching time (dashed red
   lines). \emph{Right}: The distribution of infall times for Leo~II based on
   selecting the $15$ subhalos that both reside nearest to Leo~II in the
   $\log(\mathcal{E}/V_{\rm max}^2)$--host-centric distance space (see
   Fig.~\ref{fig:bedist}) and have the same directionality in their radial
   velocity. The $t_{\rm infall}$ KDE (magenta line) is inferred using a
   gaussian kernel with a bandwidth selected via a leave-one-out cross
   validation grid search.  We adopt the peak in the distribution (black dashed
   line) as the characteristic infall time for Leo~II. The accompanying magenta
   shaded region illustrates the bounds on the uncertainty in the inferred
   $t_{\rm infall}$, corresponding to the most probable region encompassing
   $68~\%$ of the total distribution. For reference, the red line shows the
   quenching time as derived from the star formation history in the \emph{left}
   panel. The typical Leo~II-like subhalo in the simulations falls onto the host
   system for the first time $\sim 2~{\rm Gyr}$ prior to the cessation of star
   formation. This relatively fast quenching timescale is consistent with
   previous statistical work and favors ram-pressure stripping as the dominant
   quenching mechanism, perhaps assisted by tidal stripping and stellar feedback
   during pericentric passage.}
 \label{fig:example}
\end{figure*}


\subsubsection{Simulations}
\label{sec:phatelvis}

To determine the infall time of each Milky Way satellite in our sample, we
utilize the Phat ELVIS (pELVIS) suite of $12$ high-resolution,
dissipationless simulations of Milky Way-like halos along with an analytic
potential set to match the observed properties of the MW disk and
bulge \citep{kelley18}.
The suite includes $12$ isolated halos simulated within high-resolution
uncontaminated volumes spanning $1 - 3$ Mpc in size and using a particle mass of
$2.92 \times 10^{4}~\msun$ and a Plummer equivalent force softening of
$\epsilon = 25$ physical parsecs.
Within the high-resolution volumes, the halo catalogs are complete down to
$\mhalo > 3 \times 10^{6}~\msun$, $\vmax > 4.5~{\rm km}~{\rm s}^{-1}$,
$\mpeak > 8 \times 10^{6}~\msun$, and $\vpeak > 5~{\rm km}~{\rm s}^{-1}$ ---
therefore more than sufficient to track the evolution and infall time of halos
hosting MW dwarfs with $\mstar > 10^{3.5}~\msun$ \citep{gk14}.
The simulations adopt a cosmological model with the following $\Lambda$CDM
parameters: $\sigma_8 = 0.815$, $\Omega_{m} = 0.3121$,
$\Omega_{\Lambda} = 0.6879$, $n_{s} = 0.9653$, and $h = 0.6751$
\citep{planck16}.
For a detailed discussion of the simulations and the impact of the host baryonic
potential on the subhalo (i.e.~satellite) population, we refer the reader to
\citet[][see also \citealt{donghia10, gk17, sawala17, graus18}]{kelley18}.

For all subhalos with $\mpeak > 10^{8}~\msun$, we determine their infall time,
defined to be the first timestep when the subhalo passes within the host virial
radius.
Some ($\sim24\%$) of the subhalos have multiple infall times, with a median
difference between first and last infall of $3.9~\gyr$.
These differences lead to a marginal change in the infall time
inference for individual MW satellites ($\sim 0.71~\gyr$), however
they do not change the qualitative results of this work.
On average, there are $250$ subhalos per host above our adopted mass limit of
$M_{\rm peak} > 10^{8}~\msun$, with the $100\%$ scatter ranging from $129$ to
$367$.
All of the subhalos from each simulation box are combined into one final catalog
when compared to the MW satellite population.

To facilitate constraining the infall time for each MW dwarf galaxy, we match
the observed $6{\rm D}$ phase-space properties of the MW satellites to subhalo
properties in the simulations.
Using the measured infall time for each subhalo, we can map the infall time
distributions to the phase-space distributions in the simulations.
A useful method for mapping these two distributions is presented in
\citet{rocha12}, where a very strong correlation is found between the
present-day binding energy of the subhalos in simulations and their
infall times. 
We use a modified version of the subhalo binding energy presented in
\citet{rocha12} in order to match the observed binding energies to the subhalo
properties, thereby constraining the possible distribution of infall times for
each MW satellite.

\subsubsection{Binding Energy}
\label{subsubsec:BE}

As shown by \citet{rocha12}, the binding energy per unit mass,
$\mathcal{E}$, of a dark matter subhalo strongly correlates with its
infall time, such that early accreted subhalos have higher binding
energies. Following \citet{rocha12}, we define the binding energy of a
subhalo (or local dwarf) as 
\begin{equation}
\mathcal{E} =  -\phi(R) - \frac{1}{2} V^{2} \; ,
\label{eqn:BE}
\end{equation}
where $\phi(R)$ is the potential of the subhalo (or MW satellite) at the
host-centric radial distance, $R$, and $V^{2}/2{\,}$ corresponds to the kinetic
energy of the subhalo (or galaxy), with $V$ representing the magnitude of the
$3{\rm D}$ velocity vector in the frame of reference of the host.

To determine the potential of each satellite galaxy and subhalo, we assume the
host mass is distributed in an NFW potential \citep{nfw97}:
%
\begin{equation}
\phi(r) = -4 \pi G \rho_{0} r_{s}^{2} \frac{{\rm ln}(1 + r/r_{s})}{r/r_{s}} \; ,
\label{eqn:potential}
\end{equation}
where $G$ is the gravitational constant, $\rho_{0}$ is the average density of
the dark matter halo, and $r_{s}$ is the scale radius. We define
$\phi(R_{0}) = 0$ at $R_{0} = 5~{\rm Mpc}$. Varying the distance where we anchor
the potential does not significantly affect the binding energy measurement or
the subsequent distribution of infall times for each MW satellite galaxy.
Adopting $R_{0} = 5~{\rm Mpc}$ ensures that $R_{0}$ is well outside the
virial radius of the host for all of cosmic time.
The simulations include hosts spanning a range of virial masses and
we use their respective concentrations as given in Table 2 of the
pELVIS halo catalogs \citep{kelley18}.
These concentrations tend to be larger than their
dark-matter-only counterparts owing to the response of the baryonic
potential. 
For the suite of $12$ hosts, the median
concentration is $c \sim 16$.

Due the variation in the host virial masses across the simulation suite, the
absolute normalization of the binding energy also varies from host to
host. Essentially, the subhalos of more massive hosts have larger binding
energies, on average, due to their deeper potentials.
To better compare the simulations to the observational data, we combine the
subhalo populations across the various simulation boxes.
We ``normalize'' the binding energy of each satellite by dividing the
measured binding energy by the present-day host maximum circular
velocity squared ($V^{2}_{\rm max}$).
For the MW dwarfs, we assume the host potential is an NFW with the following
properties: $\mvir = 1.3 \times 10^{12}~\msun$, $\rvir = 300~{\rm kpc}$, and
$\rscale = 18.75~{\rm kpc}$ \citep{bhg16}.

\subsubsection{Subhalo Matching}
\label{subsubsec:subhalomatch}

Both the binding energy, as defined in Equation~\ref{eqn:BE}, and the radial
distance from the host are scalar quantities. 
By including the vector information (namely the direction of the
radial velocity) associated with the satellite velocities, we
endeavor to further constrain the inferred infall times using a two-step subhalo
selection criteria -- essentially a vector-based selection followed by a
scalar-based selection.
First, we select all subhalos with radial velocity components in the same
direction as the dwarf galaxy in question (e.g.~since Leo~II has a positive
radial velocity, we first select all of the subhalos that also have positive
radial velocities).
From the remaining subhalos, we select the $15$ nearest to the Milky Way dwarf
in the $\log(\mathcal{E}/V^{2}_{\rm max})$-versus-$R/R_{\rm vir}$ space (see
Figure~\ref{fig:bedist}).

For each sample of $N=15$ subhalos, we infer the underlying distribution of
infall times through kernel density estimation (KDE).  We adopt a gaussian
kernel and select the bandwidth using a grid search with leave-one-out cross
validation for each individual galaxy \citep[e.g.][]{efron82}.
The peak in each KDE is adopted as the characteristic infall time for each MW
dwarf galaxy.
To estimate the uncertainty in the inferred infall time, we integrate the KDE,
centered at the peak, until the bounds include $68\%$ of the area under the
curve, thereby approximating the $1\sigma$ width of the distribution (see the
magenta shaded region in Fig.~\ref{fig:example} as an example).

We check the validity of this method by drawing a ``test'' sample
from the simulation data and attempt to infer the true infall times of
this subset using the methodology described above. 
By construction, the total sample includes the test subhalos, such that adopting
a comparison set of $N=1$ (versus $N=15$) returns the exact subhalos in question
and a ``perfect'' infall time inference.
We estimate the accuracy of our methodology from the difference between the
inferred infall time and the true infall time for the subhalo population.
The distribution of differences in the infall time is sharply peaked at $0~\gyr$
with a standard deviation of $\sim1.0~\gyr$, suggesting that we are able to
accurately constrain the true infall time for the majority of subhalos.
It should be noted that exclusion of the vector information from our method
increases the standard deviation to $\sim3.0~\gyr$, highlighting how using both
the vector and scalar selections dramatically improves the accuracy of the
inference.
For each subhalo (or satellite), the resulting infall time distribution does not
depend strongly on the choice of $N=15$ for $5\lt N\lt 20$. We select $N=15$ so
as to balance properly sampling the underlying distribution against sampling too
much of the $\log(\mathcal{E}/V^{2}_{\rm max})$-versus-$R/R_{\rm vir}$ parameter
space.


\begin{figure*}
 \centering
 \hspace*{-0.3in}
 \includegraphics[width=5in]{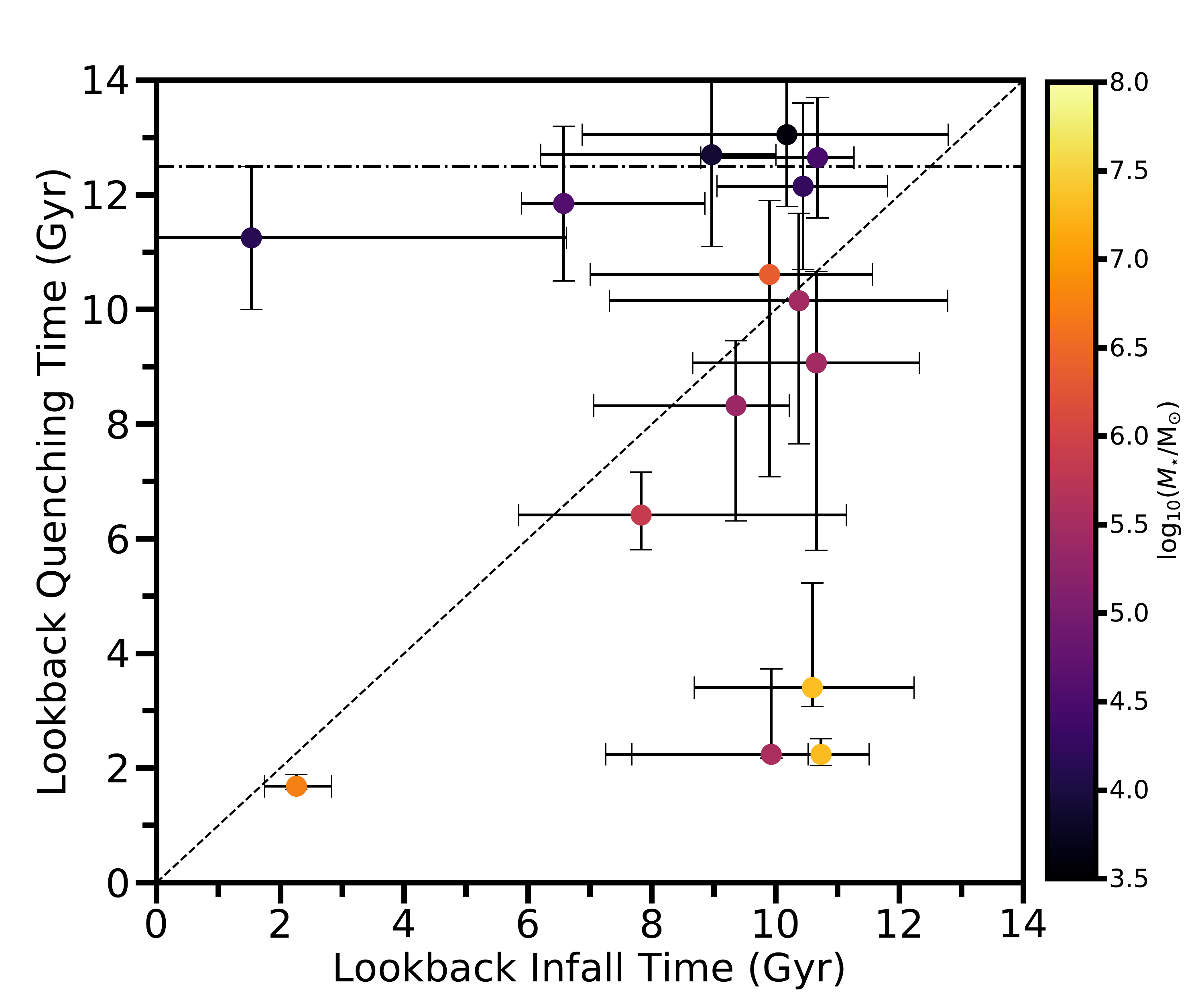}
 \caption{The relationship between quenching time ($t_{\rm quench}$) and infall
   time ($t_{\rm infall}$) for our primary sample of MW satellites. The points
   are color-coded by their present-day stellar mass, assuming $M/L_{V} = 1$
   (see color bar for scale). The low-mass ($\mstar < 10^{5}~\msun$) satellites
   have quenching times consistent with suppression of star formation due to
   cosmic reionization, as illustrated by the dot-dashed line. For the
   ``classical'' ($\mstar = 10^{5-8}~\msun$) satellites, we generally find
   quenching and infall times that are consistent with a rapid cessation of star
   formation following accretion onto the MW. The lack of satellites
   with recent infall times is largely a consequence of observational
   bias, particularly at low stellar masses. Objects that recently
   fell onto the MW system preferentially reside at larger Galactocentric
   distances, such that we are likely ``missing'' a significant number
   of low-mass satellites for which we would infer more recent infall times.
   Finally, none of the satellites studied have quenching and infall
   times that strongly favor self-quenching via internal processes.}
 \label{fig:infallvst90}
\end{figure*}


\subsection{Quenching Timescale}
As discussed previously, we define the quenching timescale ($\tau_{\rm quench}$)
as the time that a satellite continues to forms stars following infall. In other
words, the quenching timescale is the difference between the infall time and the
quenching time:
\begin{equation}
 \tau_{\rm quench} =  t_{\rm infall} - t_{\rm quench} \; ,
\label{eqn:tauq}
\end{equation}
where $t_{\rm infall}$ and $t_{\rm quench}$ are the lookback time to infall and
quenching, respectively.
In Table~\ref{table:MWdwarfs}, we report the measured infall time for
all dwarfs in the \citet{fritz18} sample within $300~\kpc$ of the MW
and the quenching time for each dwarf in our primary sample.
The uncertainty in both the infall time and quenching time are added in
quadrature to compute the total uncertainty in the quenching timescale.


\begin{figure*}
 \centering
 \hspace*{-0.3in}
 \includegraphics[width=0.8\textwidth]{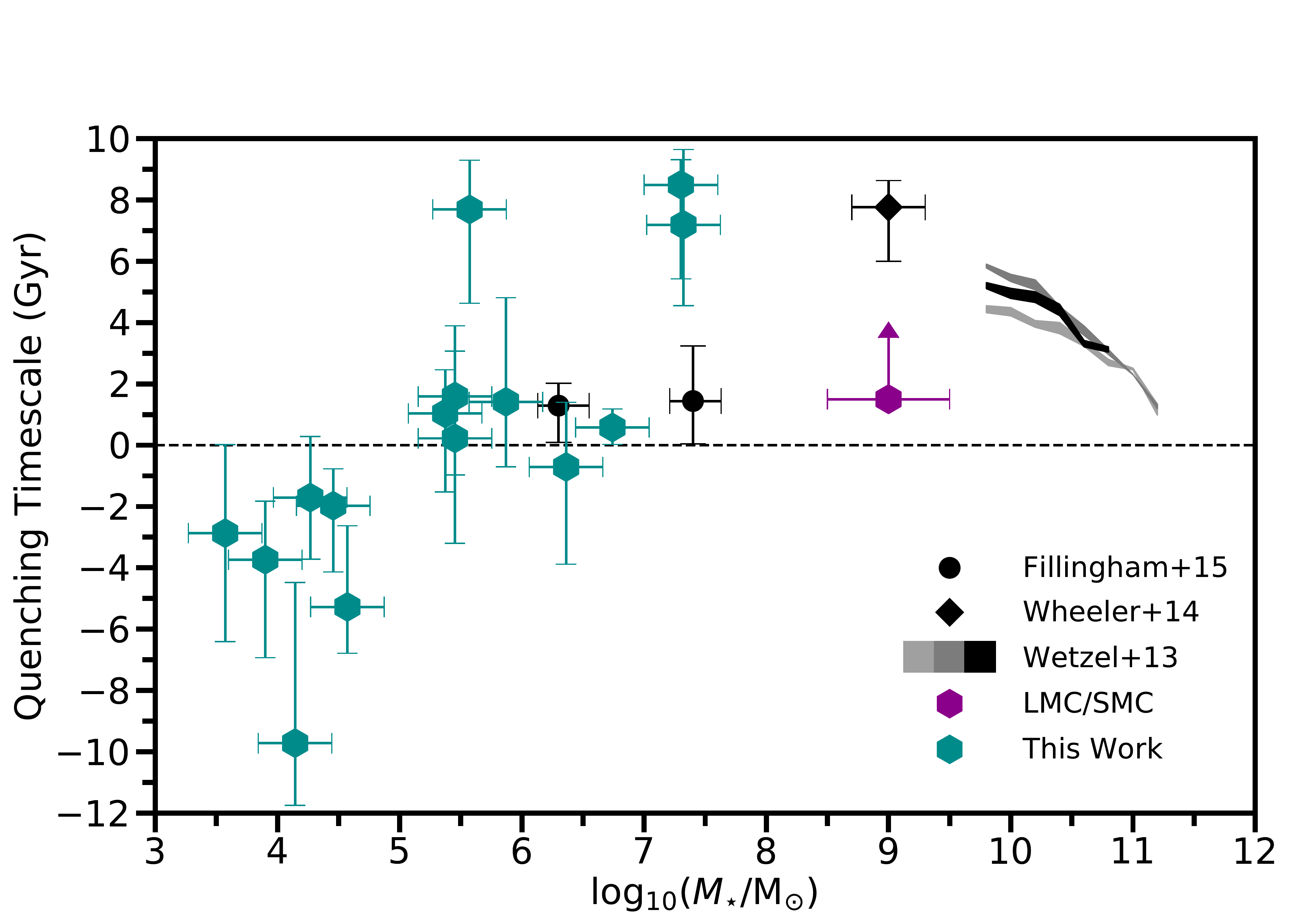}
 \caption{The quenching timescale as a function of satellite stellar mass for
   our primary sample of MW dwarfs (cyan hexagons).
   We include an estimate of the quenching timescale for both the SMC
   and LMC (magenta hexagon) based on a lookback time to first infall of
   $1.5~\gyr$ from \citet{nk13}. The black data
   points and grey bands illustrate the results of statistical studies of
   satellites in groups and clusters \citep[including the Local
   Group,][]{wetzel13, wheeler14, fham15}, with the grey bands corresponding to
   groups with estimated halo masses of $10^{12-13}$, $10^{13-14}$, and
   $10^{14-15}~\msun$ (black, dark grey, and light grey, respectively). Our
   measured quenching timescales, computed on an object-by-object basis, are in
   general agreement with the results of statistical studies, such that
   satellites with $10^{5} \lesssim \mstar/\msun \lesssim 10^{8}$ have positive
   and short quenching timescales consistent with relatively rapid environmental
   quenching. Below this mass scale ($\mstar \lesssim 10^{5}~\msun$), our
   measured quenching timescales are systematically negative, consistent with a
   scenario in which the lowest-mass galaxies quench prior to infall (i.e.~at
   the time of, or shortly following, reionization).}
 \label{fig:tqmstar-punchline}
\end{figure*}


\section{Results}
\label{sec:results}

Figure~\ref{fig:infallvst90} shows the quenching time as a function of infall
time, color-coded by stellar mass, for our primary sample of Milky Way
satellites with published {\it HST}-based star-formation histories.
For the lowest mass systems ($\mstar \lesssim 10^{5}~\msun$), we find that the
quenching time universally exceeds the infall time
($t_{\rm quench} > t_{\rm infall}$), such that the suppression of star formation
preceded infall onto the Milky Way.
In constrast, at higher satellite masses, quenching is found to occur at (or
after) the time of infall ($t_{\rm quench} \lesssim t_{\rm infall}$). 
In this manner, the quenching time--infall time space can be used to determine
the class of quenching mechanisms potentially responsible for shutting down star
formation.
Specifically, objects that fall below and to the right of the dashed line in
Figure~\ref{fig:infallvst90} are consistent with environmental quenching, where
star formation is quenched after the galaxy becomes a satellite.
The distance from the dashed line in combination with how early (or late) infall
time occurred, can further constrain which environmental quenching mechanisms
are most likely responsible.
The horizontal dot-dashed line at $t_{\rm quench} = 12.5~{\rm Gyr}$ ($z \sim 6$)
marks the approximate location of the end of reionization \citep[e.g.][]{fan06,
  robertson15}. Galaxies with $t_{\rm quench}$ consistent with this value are
potentially quenched during (or shortly after) the epoch of reionization.
The remaining region in the $t_{\rm quench}$-$t_{\rm infall}$ space, above and
to the left of the dashed line but below the dot-dashed line, favors quenching
via feedback or some other internal process such that star formation is
suppressed prior to infall.  
As evident in Fig.~\ref{fig:infallvst90}, this region of parameter space is
largely unoccupied, suggesting that internal processes are not a primary driver
of quenching within the low-mass galaxy population.
This is consistent with observations of the field population, which is
dominated by star-forming systems at these mass scales \citep{geha12}. 
Our measurements are very consistent with rapid quenching ($\tau_{\rm
  quench} \sim 1~{\rm Gyr}$) within the MW halo at a $< 2\sigma$ level.

Subtracting the quenching time from the infall time, we determine the quenching
timescale ($\tau_{\rm quench}$) for every satellite galaxy in our primary sample
(see Equation~\ref{eqn:tauq}).
In Figure~\ref{fig:tqmstar-punchline}, we plot this quenching timescale, for each
system, as a function of its stellar mass (cyan hexagons).
In addition, we include an estimate of the quenching timescale for the Small and
Large Magellanic Clouds (SMC/LMC) as denoted by the the magenta hexagon.
Under the assumption that the SMC and LMC are on first infall, orbital modeling
can estimate the lookback time at which they crossed the MW virial radius
\citep{besla07, nk13}.
We estimate the infall time for the Magellanic system to be $\sim
1.5~\gyr$ \citep[Figure 13 of][]{nk13}, which leads to a conservative
lower limit on the quenching timescale based on the framework
described in this work. 
For comparison to the estimates of the quenching timescale for individual MW
satellites, we also include the results of several statistical studies that
provide estimates of the typical quenching timescale for various satellite and
host samples in the local Universe \citep{wetzel13, wheeler14, fham15}.
Overall, the individual measures of $\tau_{\rm quench}$ for the MW satellites
are in good agreement with the results based on the analysis of ensembles.
At $\mstar \gtrsim 10^{5}~\msun$, the positive quenching timescales are
broadly consistent with the expectations from \citet{fham15}.
Meanwhile, the universally negative quenching timescales for the lowest mass
systems ($\mstar \lesssim 10^{5}~\msun$) are consistent with quenching driven
by reionization \citep{wimberly19}.
%

\section{Discussion}
\label{sec:disc}

\subsection{Quenching in Classical Dwarf Satellites}
\label{subsec:highmass}

For decades now, the suppression of star formation in the ``classical'' dwarf
galaxy population orbiting the MW ($\mstar \sim 10^{5-8}~\msun$) has been
commonly attributed to processes associated with environment
\citep[e.g.][]{lin83, blitz00}.
More recent modeling work shows that the timescale, over which the transition
from star forming to quiescent must occur (following infall onto the host), is
very short \citep{slater14, wetzel15a, fham15, simpson18}.
This short quenching timescale is consistent with ram-pressure stripping being
the dominant quenching mechanism.
In an effort to explicitly test the validity of this hypothesis,
\citet{fham16} find that gas stripping processes can reproduce the necessary
timescales if the density of the MW circumgalactic medium (CGM) is clumpy
\citep[see][for a contrasting view]{emerick16}, such that orbiting satellite
galaxies will experience elevated ram-pressure stripping when they interact with
these overdensities.
The majority of classical MW satellite galaxies in this study are consistent
with having their star formation shut down very rapidly after falling onto the
MW system and becoming a satellite (i.e.
$\tau_{\rm quench} \sim 1-2~{\rm Gyr}$).
These objects fit nicely into the environmental quenching picture put forward in
\citet{fham15,fham16}.

As evident in Figure~\ref{fig:tqmstar-punchline}, however, there are three
dwarfs (Sagittarius, Fornax, and Carina) with inferred quenching timescales
significantly longer than the short quenching timescales inferred from previous
work.
Closer inspection of Sag~I suggests that its infall time (and thus quenching
timescale) is not reliably measured via comparison to subhalos in the pELVIS
simulation suite.
As readily seen in Figure~\ref{fig:bedist}, Sag~I occupies a region of the binding
energy-vs-distance parameter space that is largely devoid of corresponding
subhalos in the simulations.
This is likely due to the fact that the pELVIS simulations do not account for
the satellite's baryonic component. In the case of a system such as Sag~I that is
being disrupted via tidal interaction with the MW, the corresponding subhalo
population in the pELVIS simulations is quickly stripped, whereas Sag~I is better
able to resist disruption due its non-negligible baryonic mass (Lazar et al., in
prep).
Put another way, at fixed host mass, low-mass subhalos are
stripped more slowly \citep{bk07, bk08}, such that at small host-centric
distance and high binding energy the subhalo population in pELVIS is
dominated by low-mass subhalos that have been preferentially accreted
at higher $z$. 
As a result, the measured infall time for Sag~I, based on comparison to pELVIS, is
biased towards early cosmic time.
Previous work modeling the associated stellar stream suggests that Sag~I was accreted onto
the MW roughly $\lesssim 2-8~{\rm Gyr}$ ago, following an orbit with a close
pericentric passage \citep[$\lesssim 20~{\rm kpc}$,][]{law10, purcell11,
  dl17}. 
If we assume $t_{\rm infall} \sim 5~{\rm Gyr}$ for Sag~I, instead of our estimate
of $t_{\rm infall} \sim 11~{\rm Gyr}$, we find a quenching timescale of
$\tau_{\rm quenching} \sim 1.6~{\rm Gyr}$, consistent with rapid suppression of
star formation following infall.

\begin{figure}
 \centering
 \hspace*{-0.2in}
 \includegraphics[width=3.2in]{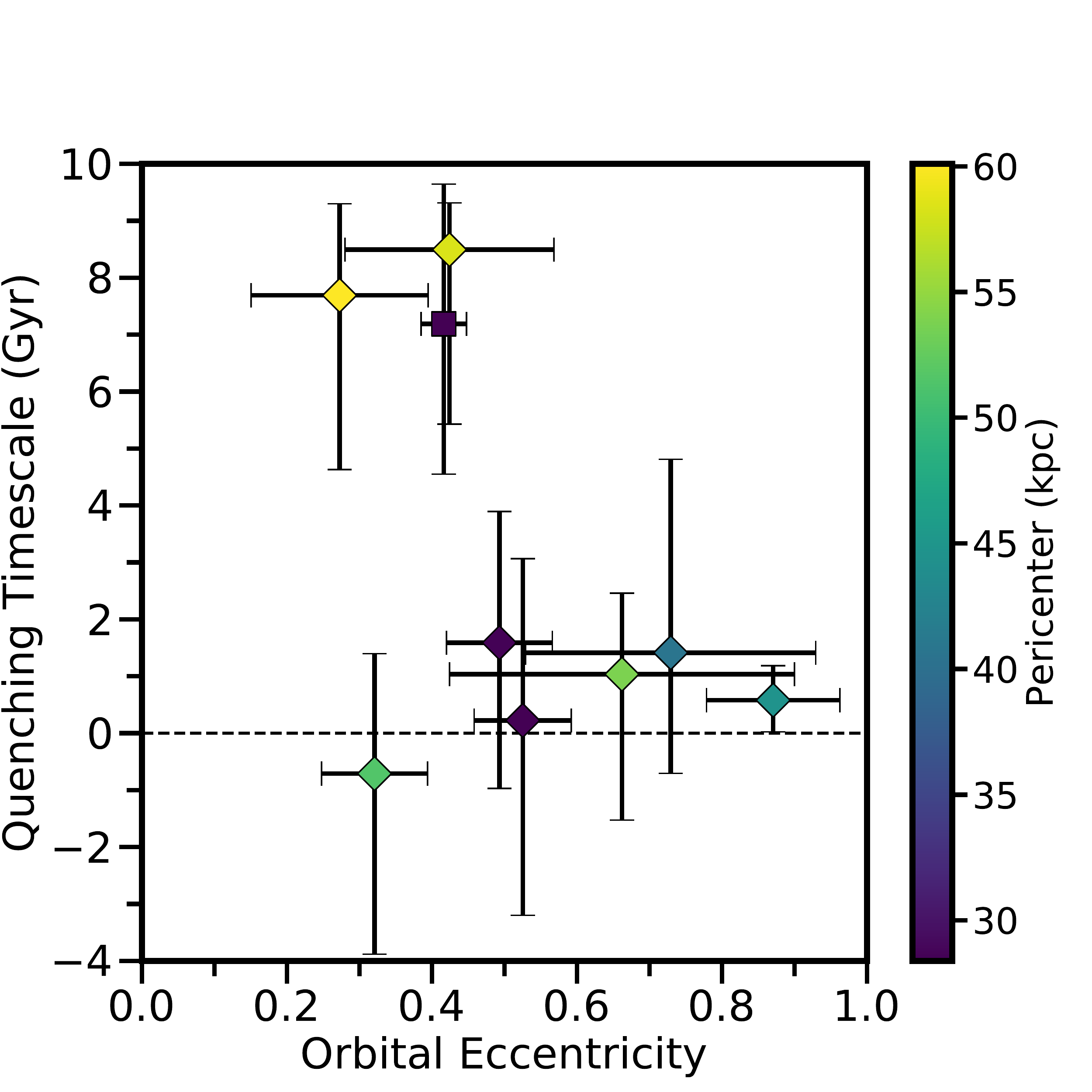}
 \caption{The quenching timescale as a function of orbital
   eccentricity, color-coded by pericentric distance, for the
   MW satellites with $\mstar > 10^{5}~\msun$ in our primary
   sample. Due to the limitations of modeling Sag using the pELVIS
   simulations, it is given a different marker (square) to distinguish
   it from the other satellites (diamonds) in this analysis.
   The satellite galaxies with the longest
   quenching timescales are on orbits with some of the smallest eccentricities --
   i.e.~more circular orbits. When combined with the relatively large
   pericentric distances for their orbits, these dwarf galaxies have likely
   experienced lower ram-pressure stripping, on average, allowing
   them to retain more of their cold gas reservoir and continue to
   form stars longer than satellites on more radial orbits. }
 \label{fig:tqecc}
\end{figure}

The other two satellites (Fornax and Carina), for which we measure
$\tau_{\rm quench} > 4~{\rm Gyr}$, suffer far less from the issues associated
with modeling Sag~I, such that the long quenching timescales inferred for these
systems are much more reliable.
These two systems, however, are outliers with regard to their orbits as inferred
by \citet{fritz18} for the ``heavy'' MW potential.
In Figure~\ref{fig:tqecc}, we plot the measured quenching timescale as a
function of orbital eccentricity, color-coded by pericentric distance,
for those satellites in our primary sample with $\mstar > 10^{5}~\msun$.
Among the massive satellites in our primary sample, Carina and Fornax
have some of the lowest orbital eccentricities (i.e.~the most circular orbits).
In addition, Carina and Fornax are on orbits with relatively large
pericenters of $60$ and $58~{\rm kpc}$, respectively.
Taken together, the orbital eccentricity and pericentric distance for Carina and
Fornax are consistent with orbits about the MW that would lead to less
efficient quenching via environmental processes (e.g.~ram-pressure stripping),
which operate with greater efficiency deeper in the host potential.
Essentially, if ram pressure is less effective at removing cold gas from an
infalling dwarf, then the system will continue to form stars for several Gyr
longer (yielding a longer $\tau_{\rm quench}$) relative to satellites of similar
mass that are on more radial orbits and experience higher ram pressure.


\subsection{Quenching in Ultra-Faint Dwarf Satellites}
\label{subsec:lowmass}

As highlighted in \S\ref{sec:intro}, the inferred SFHs of the ultra-faint dwarf
galaxy population around the MW are universally consistent with a cessation of
star formation at very early times \citep[see also][]{martin16, martin17}.
Recent work shows that these early quenching times are consistent with
reionization being responsible for the suppression of star formation in the
lowest mass galaxies \citep{brown14, wimberly19}.
This is in good agreement with hydrodynamic simulations of isolated dwarfs,
which find that reionization is effective in shutting down star formation on the
smallest mass scales \citep{bl15, fitts17, jeon17, wheeler18}.

Our work adds further evidence to this reionization quenching scenario by
determining the infall time for each of the UFDs with a measured SFH.
Not only did star formation terminate at very early times, the inferred infall times
for the UFDs occur after the SFH-based quenching time, strongly suggesting that
quenching mechanisms associated with the MW environment are not the dominant
mode of star formation cessation.
In agreement with simulations and related work, our analysis adds additional
evidence for a stellar mass scale of $\mstar \sim 10^{5}~\msun$, below which
star formation is shut down by reionization.

\subsection{Comparison to Previous Studies} 
\label{subsec:rocha}

Utilizing the recently-measured $6$D phase-space information for the MW
satellites based on \emph{Gaia} DR2 proper motions, we infer the infall time
for $37$ MW satellites through comparison to the phase-space distributions of
subhalos in the pELVIS simulation suite.
Previous work demonstrated the viability of this exercise using the Via Lactea
simulations \citep{diemand07, diemand08} and showed the existence of a strong
correlation between the binding energy and infall time for subhalos
\citep{rocha12}.
By applying a similar method to higher-resolution simulations that include a disk
potential, we recover the same qualitative trends as those found by
\citet{rocha12}. However, the normalization, slope, and scatter of the
correlation between binding energy and infall time varies from host to host. 
%
%
%
%
Additionally, previous work utilized only $4$ components of the phase space for
the majority of the MW dwarf galaxies and demonstrated that incorporating the
full $6$ components led to a stronger constraint on the infall time distribution
\citep{rocha12}.

In our analysis, we are able to take advantage of the full phase-space
measurements now available and a suite of simulations that better captures the
host-to-host scatter in the properties of the subhalo population.
Where applicable, the infall times in \citet{rocha12} and in this work agree
within $2\sigma$. However, we generally infer earlier infall times.
This difference is driven by the inclusion of the full $6$D phase-space
measurements based on \emph{Gaia} proper motions.
If we instead infer the infall time using only the Galactocentric position and
radial velocity, then we find much better agreement (within $1\sigma$) with
\citet{rocha12}.
This agreement, however, does not imply accuracy, as the overall accuracy of this
method decreases if we omit the tangential velocity component.
When compared to the true infall times for subhalos in pELVIS, the inclusion of
tangential motion to the estimate of $t_{\rm infall}$ reduces the catastrophic
failure rate from $\sim 30\%$ to $\sim 20\%$, where failure is defined to be
$|{\Delta t_{\rm infall}}| / (1 + t_{\rm infall, true}) > 0.2$.  
Thus, when possible, the entire $6$D phase space should be used to infer the
most likely infall time for the MW dwarf galaxies.

\citet{weisz15} also characterize the quenching time for the MW
satellite galaxies using \emph{HST}-based star-formation histories.
That work, however, was based on the less accurate infall times from
\citet{rocha12}, leading to the conclusion that many of the MW satellite
galaxies with $\mstar \gtrsim 10^{5}~\msun$ had their star formation shut down
prior to becoming a satellite, in conflict with the established picture of
environmental quenching.
With the updated phase-space measurements from \emph{Gaia} DR2, the infall times
for these systems are now better constrained and favor a scenario in which
environmental quenching dominates at $\mstar \sim 10^{5-8}~\msun$.

\section{Summary}
\label{sec:summary}

Our work highlights the power of combining detailed observations of galaxies in
the nearby Universe with cosmological simulations.
We characterize the infall time for the population of MW satellite galaxies with
\emph{Gaia} DR2 based proper motion measurements from \citet{fritz18}.
When combined with quenching times inferred from \emph{HST}-based SFHs, we
measure the quenching timescale for individual satellites, facilitating a
detailed, object-by-object study of satellite quenching. 
The principal results of our analysis are as follows:

\begin{itemize}[leftmargin=0.2cm]

\item Using the Phat ELVIS suite of high-resolution, zoom-in
  simulations, we develop a mapping from binding energy to subhalo
  infall time that allows for relatively precise constraints on the
  infall time of MW satellites. As shown in
  Table~\ref{table:MWdwarfs}, using this mapping, we estimate the
  infall time for all MW satellites with a {\it Gaia} DR2 proper
  motion measurement from \citet{fritz18} and currently reside within
  $300~\kpc$ of the MW. \\ 

\item The inferred quenching timescales for satellites of the Milky Way with
  $\mstar = 10^{5-8}~\msun$ is broadly consistent with the rapid cessation of star
  formation following infall, supporting the model for satellite quenching
  developed by \citet{fham15, fham16, fham18}. \\

\item The infall and quenching times measured for the UFD satellites of the MW
  are consistent with quenching via reionization and support a critical scale of
  $\mstar \sim 10^{5}~\msun$ below which reionization is effective in
  suppressing star formation at early cosmic time. \\ 

\item Within the sample of MW satellites studied (at $\mstar < 10^{8}~\msun$),
  our measurements of infall and quenching times are fully consistent with
  quenching via reionization or environment, such that there is no
  need for satellites at these mass scales to be quenched due to
  internal processes -- e.g.~self-quenching via feedback.

\end{itemize}

\section*{acknowledgements} 
We thank Josh Simon and Alex Ji for useful conversations that helped shape
some of this analysis.
%
%
This work was supported in part by NSF grants AST-1815475, AST-1518257,
AST-1517226, AST-1009973, and AST-1009999.
Additional support was provided by NASA through grants AR-12836, AR-13242,
AR-13888, AR-13896, GO-14191, and AR-14289 from the Space Telescope Science
Institute, which is operated by the Association of Universities for Research in
Astronomy, Inc., under NASA contract NAS 5-26555.
MBK also acknowledges support from NSF CAREER grant AST-1752913 and NASA grant
NNX17AG29G.
Support for SGK was provided by an Alfred P. Sloan Research
Fellowship, NSF Collaborative Research Grant $\#$1715847 and CAREER grant
$\#$1455342, and NASA grants NNX15AT06G, JPL 1589742, 17-ATP17-0214
MSP acknowledges funding by NASA through Hubble Fellowship grant
$\#$HST-HF2-51379.001-A awarded by the Space Telescope Science
Institute, which is operated by the Association of Universities for
Research in Astronomy, Inc., for NASA, under contract NAS5-26555.
CW was supported by the Lee A.~DuBridge Postdoctoral Scholarship in
Astrophysics. 

This research has made use of NASA's Astrophysics Data System
Bibliographic Services. 
This research also utilized {\texttt{Astropy}}, a community-developed
core Python package for Astronomy \citep{astropy13}.
Additionally, the Python packages {\texttt{NumPy}} \citep{numpy},
{\texttt{iPython}} \citep{ipython}, {\texttt{SciPy}} \citep{scipy},
{\texttt{matplotlib}} \citep{matplotlib}, and {\texttt{scikit-learn}}
\citep{sklearn} were utilized for our data analysis and presentation.

\bibliography{gaiaquench}

\appendix

\section{Infall Time Distributions}
\label{app:infall}

For completeness, we show the infall time distributions for every galaxy in our
primary sample. See \S\ref{subsubsec:subhalomatch} for a description of how the
observed properties of the MW satellites are matched to the subhalo population
in the Phat ELVIS simulation suite, in order to infer the infall time
for each MW satellite in our primary sample.


\begin{figure*}
 \centering
 \hspace*{-0.3in}
 \includegraphics[width=7.0in]{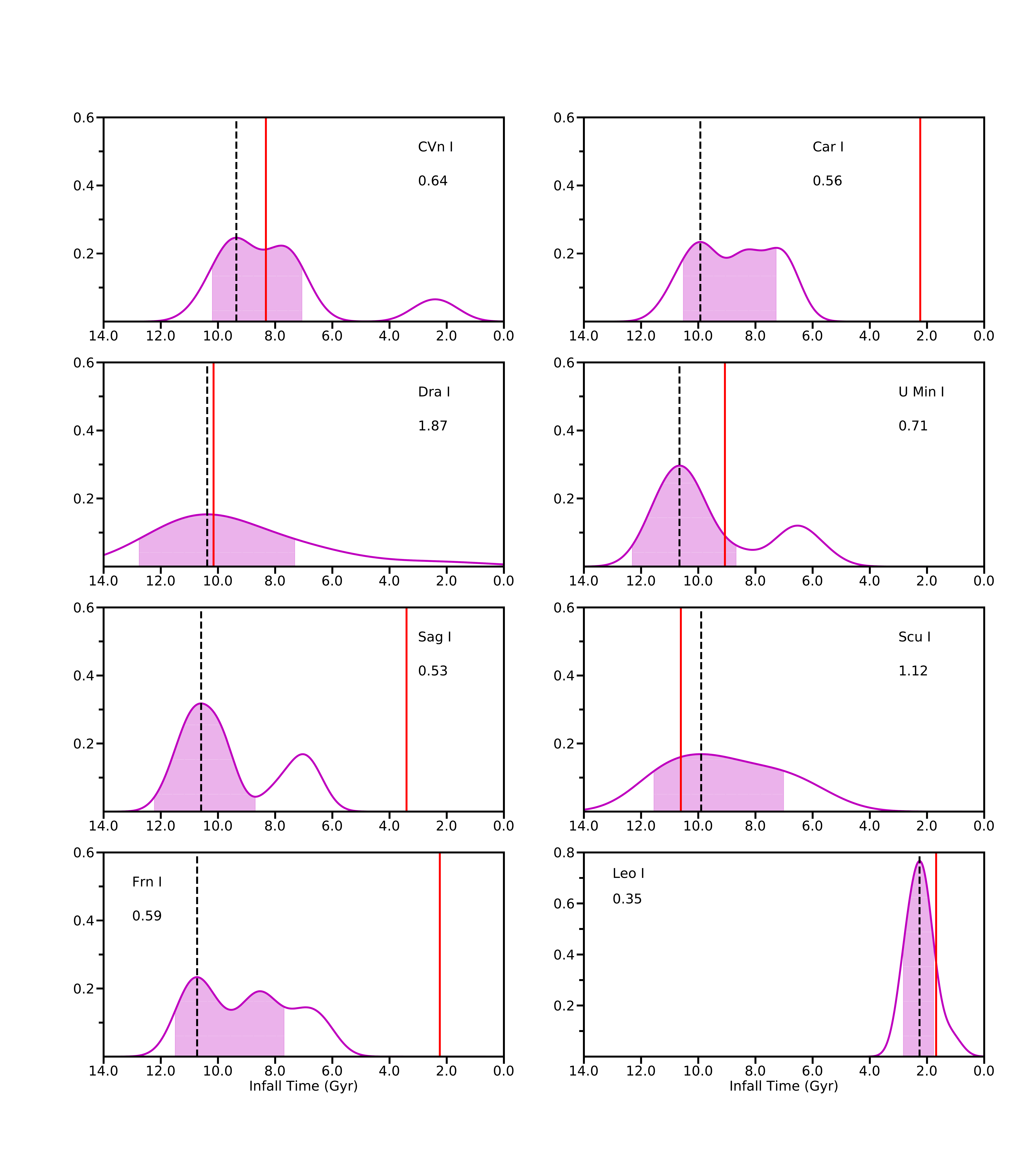}
 \caption{Same as the \emph{right} panel of Figure~\ref{fig:example} for the
   remaining ``classical'' MW satellite galaxies.  The distribution of infall
   times for each respective MW satellite based on selecting the $15$ closest
   subhalos in the $\log(\mathcal{E}/V_{\rm max}^2)$--host-centric distance
   space (see Fig.~\ref{fig:bedist}) and the same directionality in their radial
   velocity.  The distribution of $t_{\rm infall}$ (magenta line) is smoothed
   with a gaussian kernel with a bandwidth chosen by a ``leave-one-out'' cross
   validation grid search, to allow for a cleaner inference of the peak of the
   distribution.  The adopted bandwidth (in Gyr) is shown below each galaxy name
   in the respective panel. We adopt the peak in the distribution (black dashed
   line) as the characteristic infall time for each MW satellite. The
   accompanying magenta shaded region illustrates the adopted uncertainty on the
   characteristic $t_{\rm infall}$, corresponding to the most probable region
   encompassing $68~\%$ of the total distribution.  For reference, the red line
   shows the quenching time as derived from the \emph{HST}-based SFH. }
 \label{fig:classicalinfalls}
\end{figure*}



\begin{figure*}
 \centering
 \hspace*{-0.3in}
 \includegraphics[width=6.5in]{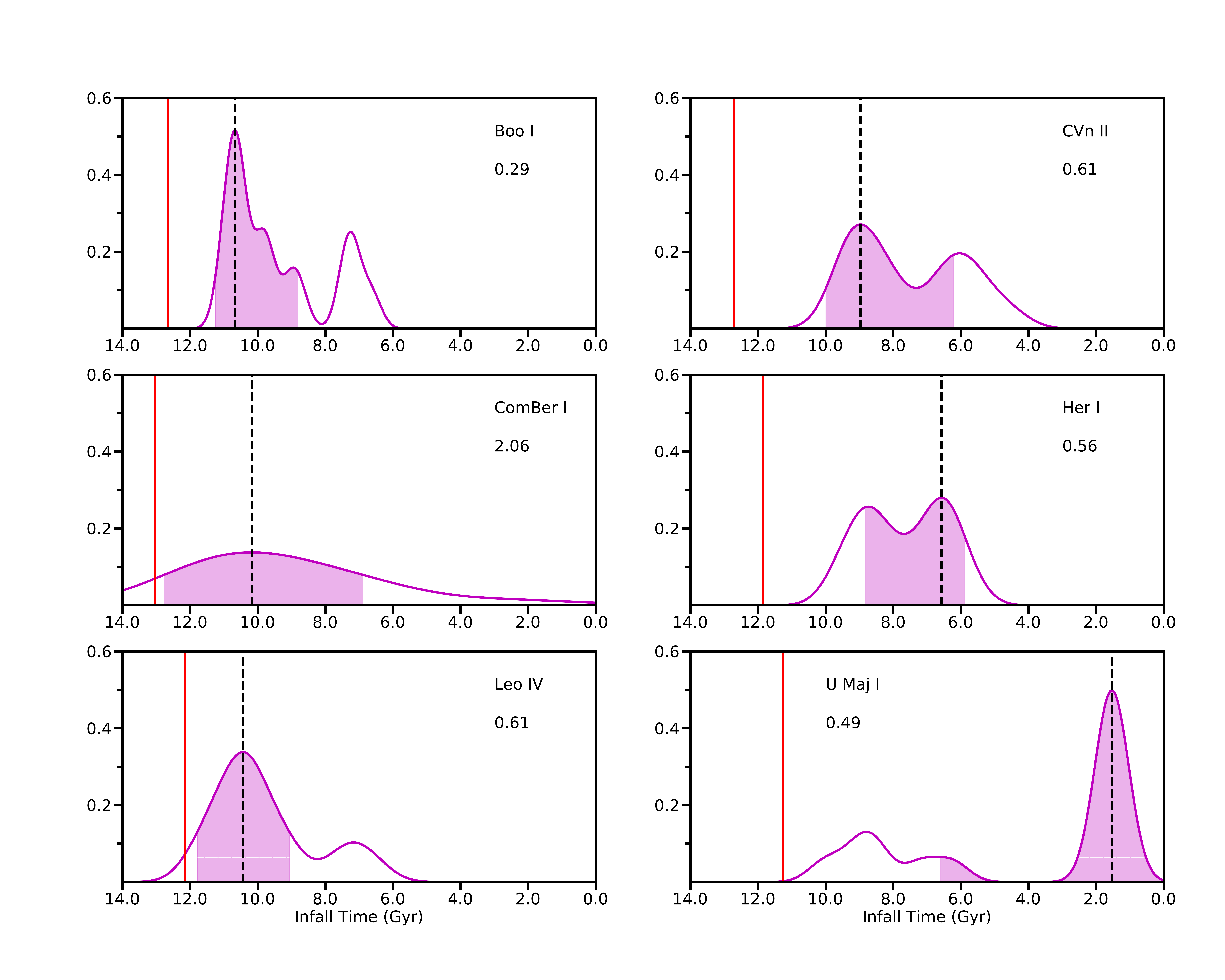}
 \caption{Same as Figure~\ref{fig:classicalinfalls}, except for the Ultra-Faint MW
   satellite galaxies (UFDs).
  }
 \label{fig:UFDinfalls}
\end{figure*}


\label{lastpage}
\end{document}